\documentclass[12pt]{article}
\usepackage{amssymb,amsmath,epsfig,graphicx}
\begin{document}
\title{\bf Modified Gauss-Bonnet Gravity with Radiating Fluids}

\author{M. Farasat Shamir\thanks{farasat.shamir@nu.edu.pk} \ and
M. Awais Sadiq \thanks{awaismalhi007@gmail.com}\\\\
Department of Sciences and Humanities, \\National University of
Computer and Emerging Sciences,\\ Lahore Campus, Pakistan.}

\date{}

\maketitle
\begin{abstract}
The main purpose of this paper is to investigates structure scalars in the context of $f(\mathcal{G}, T)$ gravity, where $\mathcal{G}$ is the Gauss-Bonnet invariant and $T$ is the trace of stress energy tensor. For this aim, we have considered the spherically symmetric spacetime and dissipative anisotropic fluid coupled with radiation and heat ejecting shearing matter distributions. We have found these scalar variables by orthogonally decomposing the Riemann curvature tensor in $f(\mathcal{G}, T)$ gravity. Moreover, the evolution equations of shear and expansion are also developed with the help of these scalar functions. We have also analyzed these scalars by taking constant $\mathcal{G}$ and $T$ for dust cloud. The physical behavior of structure scalars for radiating matter distributions has been examined in the presence of modified gravity. It is shown that the evolutionary stages of relativistic stellar structures can be explored via modified scalar functions.

\end{abstract}


\section{Introduction}
According to recent observations, the accelerating expansion of the universe is one of the most astounding discovery in the generation of astrophysics. Relativistic consequences need to be taken into consideration in the investigation of the gravitational stellar system. Vivid examples of celestial gravitational system are black holes, neutron stars, quark stars, white dwarfs in which these impacts have vital results. So it becomes important to take feasible gravitational theories into account to study these systems. The results originating from Ia Supernova and cosmic microwave background (CMB) radiation, etc. \cite{piet1} have made a remarkable revolution in the area of General Relativity, therefore opening a platform for research. These observations show that there is an accelerating expansion in our universe. According to the recent investigations, the outcomes developing from the Planck satellite \cite{plan1}-\cite{plan3}, the BICEP2 experiment \cite{bice1}-\cite{bice3}, the Wilkinson Microwave anisotropy probe (WMAP) \cite{koma1}-\cite{hins1} and Sloan Digital Sky Surveys (SDSS) \cite{tegm1}, it turns out that 68\% of universe is composed of dark energy, 27\% is the dark matter and the rest is ordinary matter which is less than 5\%. This has inspired many researchers to discover the mysterious nature of dark energy that is assumed to pass through the entire space with due course of expansion. Dark energy holds a huge amount of negative stress with repulsiveness but its essential capabilities are still not recognised. So as to contemplate the idea of dark energy, modified gravitational theories are introduced after generalizing the Einstein Hilbert action. These modified gravitational theories act as a replacement to solve the unsolved mystery of the universe which is believed to be the actual reason of the expansion in our cosmos. Nojiri and Odintsov \cite{noji1} explained that how these modified gravitational theories are important in investigating the evolutionary phases of the universe. Modified gravity theories like $f(R)$, $f(R, T)$, $f(\mathcal{G})$ and $f(\mathcal{G}, T)$ etc., have been constructed after being motivated by the first hypothesis \cite{azad1}-\cite{shar1}. One of the most talked about theory is $f(R)$ theory of gravity acquired by supplanting the Ricci scalar with $f(R)$ in Einstein Hilbert action. This theory was gotten after quite a while from the appearance of the Einstein's relativity to separate conceivable options. Furthermore it was concentrated quite often by specialists to renormalize general relativity which demands higher curvature dark source variables in the Einstein Hilbert action \cite{eddi1}-\cite{utiy1}. This theory pulled in numerous astrophysicists in achieving elucidation of cosmic expansion on account of the quadratic Ricci scalar corrections \cite{star2}. These modified gravitational theories are built up by including or substituting the curvature scalars, topological properties and their derivatives in Einstein Hilbert action.

Another modified gravity that has picked up prominence over the most recent couple of years is Gauss-Bonnet gravity generally known as $f(\mathcal{G})$ gravity where $\mathcal{G}$ is the Gauss-Bonnet invariant \cite{cogn1}-\cite{noji4}. Cosmic acceleration expansion may be observed in $f(\mathcal{G})$ gravity due to the presence of de-Sitter point. The outstanding quality of this theory is that the presence of Gauss-Bonnet term may avert vague involvements and uniforms the gravitational activity \cite{chib1}. Further, $f(R, \mathcal{G})$ theory was introduced \cite{noji5} by presenting Ricci scalar alongside Gauss-Bonnet invariant. In similar way, $f(\mathcal{G}, T)$ theory was developed by Sharif and Ikram \cite{shar1} that contains the Gauss-Bonnet invariant and the trace of energy momentum tensor. Some fascinating work has been done in the later past utilizing modified Gauss-Bonnet theories.

Shamir and Ahmad \cite{sham1} developed some cosmological feasible $f(\mathcal{G}, T)$ gravity models by using Noether symmetry approach and determined some exact solutions for flat FRW universe with $f(\mathcal{G}, T)$ corrections. Sharif and Ikram \cite{shar3} recreated some known cosmological solutions in $f(\mathcal{G}, T)$ gravity and studied the stability of specific models with linear perturbations in FRW universe. The same authors \cite{shar4} investigated the stability of Einstein universe through linear perturbations in $f(\mathcal{G}, T)$ gravity for conserved and non conserved stress energy tensor and deduced that if the model parameters are picked appropriately then Einstein universe exists for all estimations of the equation of state parameter. Shamir and Ahmad \cite{sham2} examined Noether symmetries of LRS Bianchi type $I$ universe and used anisotropic effect to discuss cosmological models in $f(\mathcal{G}, T)$ gravity. Shamir \cite{sham3} investigated the exact solutions of field equations in $f(\mathcal{G}, T)$ gravity for LRS Bianchi type $I$ space-time with anisotropic background and found cosmological solutions of field equations. Recently, Bhatti and his collaborator \cite{bhat1} studied the evolution of compact stars in $f(\mathcal{G}, T)$ theory of gravity with some specific $f(\mathcal{G}, T)$ models and concluded that the compactness of the star increases for different models of modified gravity.

Some interesting work has been done in exploring the effects of anisotropic pressure in the investigation of structure formalism of the stellar frameworks and their development. Bower and Liang \cite{bowe1} investigated the stability of locally anisotropic fluid configurations to study the impacts of radial and tangential pressure variables. Chan et al. \cite{chan1} explored the impacts of anisotropy of the development of dynamic instabilities of relativistic spherical matter configurations and derived that the presence of anisotropy in the stellar framework has significantly changed the structures of celestial system. Chakraborty et al. \cite{chak1} examined pressure anisotropy distributions on quasi spherical model and concluded that such configurations of pressure could deter the occurrence of naked singularity. Hillebrandt and Steinmetz \cite{hill1} examined the numerical method to explore the stable schemes of anisotropic of compact stars and discovered some stability regions in resemblance with isotropic models. Herrera and Santos \cite{herr4} analysed the exposure of locally anisotropic pressure and its effects on self gravitating systems. Bamba et al. \cite{bamb1} investigated the dynamics of collapsing stellar system and found that $R^\alpha$ corrections provide a cosmological singularity free model. Herrera et al. \cite{herr1} studied the stability of shear free condition for spherically symmetric anisotropic fluid distributions and found that locally anisotropic pressure and density inhomogeneity have an effect on the stability of shear free condition. Sharif and Yousaf \cite{shar5} found some exact systematic models for the spherical symmetry anisotropic structure with the effect of shear free condition. Tewari et al. \cite{tewa1} studied anisotropic fluid distributions for a spherically symmetric model and presented some relativistic models that could be useful to comprehend different characteristics of compact star models.

Structural variables, such as energy density, locally anisotropic pressure and Weyl tensor, etc,. can be used to explore the evolutionary development of stellar models. A framework starts collapsing once it encounters an inhomogeneous stellar state. The anisotropy and irregularities in the energy density possess significant role in the collapsing systems and consequently in creating theory of relativistic stellar structure development. Penrose and Hawking \cite{penr1} investigated energy density irregularities of spherical relativistic stars through Weyl scalar. Herrera et al. \cite{herr5} discussed energy density inhomogeneity and local anisotropy for spherical compact stellar object. Herrera et al. \cite{herr3} discussed the gravitational arrow of time by relating Weyl scalar with energy density inhomogeneity and anisotropic pressure. In another context, Herrera and his collaborators \cite{herr2} examined the structure and evolution of compact stars with the help of some structure scalars obtained from the orthogonal splitting of curvature tensor. Sharif et al. \cite{shar27} explored the role of structure scalars for cylindrical self gravitating systems by taking $f(R)$ into account. Sharif and Yousaf \cite{shar6} described the stability of the energy density in matter fluid configurations by considering three parametric models in Palatini $f(R)$ gravity. Yousaf et al. \cite{yous1} explored the evolutionary phases of stellar systems in $f(R, T)$ theory of gravity through structure scalars.

The main focus of this work is to investigate the influence of $f(\mathcal{G}, T)$ gravity in the construction of structure scalars. Furthermore, we examine the role of these scalar variables in the expansion and evolution equations for dissipative spherical distributions of anisotropic stellar systems. The format of this paper is as follows: In section \textbf{2} we derive the field equations of modified gravity with anisotropic fluid distributions and then link the Weyl scalar with structural variables. In section \textbf{3} modified scalar functions are constructed and their role in the study of self gravitating systems is discussed. Section \textbf{4} demonstrates the role of these scalar functions for relativistic dust cloud with constant $\mathcal{G}$ and $T$. The main outcomes are discussed in the last section.

\section{Modified Field Equations}
The general action of $f(\mathcal{G},T)$ is given by \cite{shar1}
\begin{equation}\label{1}
\mathcal{S}_{f(\mathcal{G},T)} = \frac{1}{2\kappa^2}\int d^4x \sqrt{-g}[R+f(\mathcal{G},T)]+\int d^4x\sqrt{-g}\mathcal{L}_m,
\end{equation}
where $T$ is the trace of energy momentum tensor, $g$ is the determinant of metric tensor, $R$ indicates the Ricci Scalar, $\mathcal{L}_M$ represents the Lagrangian coupled with matter and $\kappa$ symbolizes the coupling constant. For straightforwardness, we take $\kappa^2=1$.

The energy momentum tensor can be determined as \cite{land1}
\begin{equation}\label{2}
T_{\alpha\beta}=-\frac{2}{\sqrt{-g}}\frac{\delta(\sqrt{-g}\mathcal{L}_m)}{\delta g^{\alpha \beta}}.
\end{equation}
If we suppose that $\mathcal{L}_M$ only depends on the components of $g_{\alpha\beta}$ but does not rely on its derivatives, then Eq.(2) generates
\begin{equation}\label{3}
T_{\alpha\beta}=g_{\alpha \beta}\mathcal{L}_m - 2\frac{\partial\mathcal{L}_m}{\partial g_{\alpha\beta}}.
\end{equation}
Varying the action (1) with respect to $g_{\alpha\beta}$, we get the following field equations of $f(\mathcal{G},T)$ gravity
\begin{equation}\label{4}
\begin{split}
R_{\alpha\beta}-\frac{1}{2}Rg_{\alpha\beta}&=T_{\alpha\beta}-(T_{\alpha\beta}+\Theta_{\alpha\beta})f_T(\mathcal{G},T) +\frac{1}{2}g_{\alpha\beta}f(\mathcal{G},T)-(2RR_{\alpha\beta} \\
& -4R^\xi_\alpha R_{\xi \beta} -4R_{\alpha\xi\beta\eta}R^{\xi\eta}+2R_{\alpha}^{\xi\eta\delta}R_{\beta\xi\eta\delta})f_\mathcal{G}(\mathcal{G},T) -(2Rg_{\alpha\beta}\nabla^2 \\
& -2R\nabla_\alpha \nabla_\beta
-4g_{\alpha\beta}R^{\xi\eta}\nabla_\xi\nabla_\eta-4R_{\alpha\beta}\nabla^2+4R^\xi_\alpha\nabla_\beta\nabla_\xi \\
& +4R^\xi_\beta\nabla_\alpha\nabla_\xi +4R_{\alpha\xi\beta\eta}\nabla^\xi\nabla^\eta)f_\mathcal{G}(\mathcal{G},T),
\end{split}
\end{equation}
 where $f_\mathcal{G}(\mathcal{G},T)=\frac{\partial f(\mathcal{G},T)}{\partial \mathcal{G}}$,
$f_T(\mathcal{G},T)=\frac{\partial f(\mathcal{G},T)}{\partial T}$, and $\nabla^2=\square=\nabla_\alpha\nabla^\alpha$ represents the d'Alembert operator. Einstein field equations are retrieved by putting $f(\mathcal{G}, T)=0$  in Eq.(4). Moreover, field equations for $f(\mathcal{G}, T)$ gravity lessens to field equations for $f\mathcal{G}$ gravity by superseding $f(\mathcal{G}, T)$ with $f(\mathcal{G})$.

The trace of Eq.(4) is given as
\begin{equation}\nonumber
\begin{split}
& R+T-(T+\Theta)f_T(\mathcal{G}, T)+2f(\mathcal{G}, T)+2\mathcal{G}f_\mathcal{G}(\mathcal{G}, T)-2R\nabla^2f_\mathcal{G}(\mathcal{G}, T) \\
& +4R_{\alpha\beta}\nabla_\alpha\nabla_\beta f_\mathcal{G}(\mathcal{G}, T)=0,
\end{split}
\end{equation}
The divergence of Eq.(4) is found as
\begin{equation}\nonumber
\begin{split}
\nabla^\alpha T_{\alpha\beta}=\frac{f_T(\mathcal{G}, T)}{1-f_T(\mathcal{G}, T)}\Bigg[(T_{\alpha\beta}+\Theta_{\alpha\beta})\nabla^\alpha(ln f_T(\mathcal{G}, T))+\nabla^\alpha\Theta_{\alpha\beta} -\frac{1}{2}g_{\alpha\beta}\nabla^\alpha T \Bigg]
\end{split}
\end{equation}
$\Theta_{\alpha\beta}$ can be obtained as follows
\begin{equation}\label{5}
\Theta_{\alpha\beta}=g^{\xi\eta}\frac{\delta T_{\xi\eta}}{\delta g_{\alpha\beta}}.
\end{equation}

Varying Eq.(3) to get a functional expression for $\Theta_{\alpha\beta}$
\begin{equation}\label{6}
\frac{\delta T_{\alpha \beta}}{\delta g^{\xi\eta}}=\frac{\delta g_{\alpha \beta}}{\delta g^{\xi\eta}}\mathcal{L}_m + g_{\alpha\beta}\frac{\partial \mathcal{L}_m }{\partial g^{\xi\eta}}-2\frac{\partial^2\mathcal{L}_m}{\partial g^{\xi \eta}\partial g^{\alpha \beta}}.
\end{equation}
Substituting Eq.(6) in (5), we get
\begin{equation}\label{7}
\Theta_{\alpha\beta}=-2T_{\alpha\beta}+g_{\alpha\beta}\mathcal{L}_m-2g^{\xi\eta}\frac{\partial^2\mathcal{L}_m}{\partial g^{\alpha\beta}\partial g^{\xi \eta}}.
\end{equation}
We consider a non-static spherically distribution of the fluid combined with locally anisotropic fluid and shear viscosity dispersing in the form of heat flow and null radiation. We assume the subsequent from of the energy momentum tensor
\begin{equation}\label{8}
T_{\alpha \beta} = P_\perp h_{\alpha \beta} + \mu V_{\alpha} V_\beta + \Pi \chi_\alpha \chi_\beta + \varepsilon \ell_\alpha \ell_\beta + q (\chi_\beta V_\alpha + \chi_\alpha V_\beta) -
2\eta \sigma_{\alpha \beta},
\end{equation}
where $P_{\perp}$ and $P_r$ is the tangential and radial pressure, $\mu$ represents the energy density, $h_{\alpha \beta}$ expresses the projection tensor, $q_\alpha$ indicates the thermal flow, $\chi_\alpha$ is the unit four vector in the radial direction, $\eta$ signifies the coefficient of shear tensor, $\sigma_{\alpha\beta}$ is the shearing viscous tensor, and $\Pi$ is the difference of $P_r$ and $P_{\perp}$, $P_r-P_{\perp}$. Eq.(7) takes the form \cite{hark1}
\begin{equation}\label{9}
\Theta_{\alpha\beta}=-2T_{\alpha\beta}-\mu g_{\alpha\beta}.
\end{equation}
The field equations of $f(\mathcal{G},T)$ gravity are given as
\begin{equation}\label{10}
G_{\alpha \beta} = T_{\alpha \beta}^{eff},
\end{equation}
where
\begin{equation}\label{11}
\begin{split}
T_{\alpha\beta}^{eff}&=(1+f_T)T_{\alpha\beta}+\mu g_{\alpha\beta}f_T +\frac{1}{2}g_{\alpha\beta}f(\mathcal{G},T)-(2RR_{\alpha\beta}
-4R^\xi_\alpha R_{\xi \beta} \\  &-4R_{\alpha\xi\beta\eta}R^{\xi\eta}+2R_{\alpha}^{\xi\eta\delta}R_{\beta\xi\eta\delta})f_\mathcal{G}(\mathcal{G},T) -(2Rg_{\alpha\beta}\nabla^2 -2R\nabla_\alpha \nabla_\beta \\
&-4g_{\alpha\beta}R^{\xi\eta}\nabla_\xi\nabla_\eta-4R_{\alpha\beta}\nabla^2+4R^\xi_\alpha\nabla_\beta\nabla_\xi
+4R^\xi_\beta\nabla_\alpha\nabla_\xi \\
& +4R_{\alpha\xi\beta\eta}\nabla^\xi\nabla^\eta)f_\mathcal{G}(\mathcal{G},T),
\end{split}
\end{equation}
which is the effective stress energy tensor comprising of matter and dark energy respectively.

We consider a non-static spherically symmetric space-time
\begin{equation}\label{12}
ds^2 = -A^2(t,r)dt^2 + H^2(t,r)dr^2 + C^2(t,r)d\theta^2 + C^2(t,r) \sin^2 \theta d\phi^2.
\end{equation}
The quantities $\ell^\alpha$ and $V^{\alpha}$ in Eq.(8) indicates null four vector and four velocity of the fluid. The four vectors $\ell^\alpha=\frac{1}{A}\delta^\alpha_0+\frac{1}{H}\delta^\alpha_1$, $\chi^\alpha$, and $V^\alpha=\frac{1}{A}\delta^\alpha_0$ satisfy
\begin{eqnarray}\nonumber 
  \chi^\alpha \chi_\alpha=1,& V^\alpha V_\alpha = -1,& \ell^\alpha \ell_\alpha=0 \\ \chi^\alpha V_\alpha=0,& V^\alpha q_\alpha = 0,& \ell^\alpha V_\alpha=-1.
\end{eqnarray}
The shear $\sigma_{\alpha \beta}$ and expansion $\Theta$ of the fluid are given by \cite{herr1}
\begin{eqnarray}
  \Theta &=& V^\alpha_{;\alpha}, \\
  \sigma_{\alpha\beta} &=& V_{(\alpha;\beta)}+a_{(\alpha}V_{\beta)}-\frac{1}{3}\Theta h_{\alpha\beta},
\end{eqnarray}
where $a_\alpha$ is the 4-acceleration and $h_{\alpha\beta}=g_{\alpha\beta}+V_{\alpha}V_{\beta}$.
The non zero components of the shear tensor are
\begin{eqnarray}\nonumber
  \sigma_{11} &=& \frac{2}{3}H^2\sigma, \\
  \sigma_{22} &=& \frac{\sigma_{33}}{sin^2\theta}=-\frac{1}{3}C^2\sigma,
\end{eqnarray}
and the scalars indicating shearing motion and expansion of the fluid are given as follows
\begin{eqnarray}
  \sigma &=& \frac{1}{A}\Bigg(\frac{\dot{H}}{H}- \frac{\dot{C}}{C}\Bigg), \\
  \Theta &=& \frac{1}{A} \Bigg(\frac{2\dot{C}}{C} + \frac{\dot{H}}{H}\Bigg).
\end{eqnarray}
where dot represents the partial derivative with respect to $t$.

The field equations for the line element (12) are as follows
\begin{equation}\label{19}
G_{00} = A^2\Bigg[\mu + \varepsilon+\varepsilon f_T - \frac{\mathcal{G}}{2}\Bigg(\frac{f}{\mathcal{G}}-f_\mathcal{G}\Bigg) - \frac{\psi_{00}}{A^2}\Bigg],
\end{equation}
\begin{equation}\label{20}
G_{01} = AH\Bigg[-(1+f_T)(q+\varepsilon)- \frac{\psi_{01}}{AH}\Bigg],
\end{equation}
\begin{equation}\label{21}
G_{11} = H^2\Bigg[\mu f_T + (1+f_T)(P_r + \varepsilon - \frac{4}{3}\eta\sigma )+\frac{\mathcal{G}}{2}\Bigg(\frac{f}{\mathcal{G}}-f_\mathcal{G}\Bigg)- \frac{\psi_{11}}{H^2}\Bigg],
\end{equation}
\begin{equation}\label{22} G_{22}= C^2\Bigg[(1+f_T)(P_\perp + \frac{2}{3}\eta\sigma) +\mu f_T + \frac{\mathcal{G}}{2}\Bigg(\frac{f}{\mathcal{G}}-f_\mathcal{G}\Bigg)-\frac{\psi_{22}}{C^2}\Bigg],
\end{equation}

The Misner-Sharp mass function is given by \cite{misn1}
\begin{equation}\label{23}
m(t,r)=\frac{C}{2} \Bigg(1+\frac{\dot{C}^2}{A^2}-\frac{C'^2}{H^2}\Bigg). \end{equation}
The four-velocity $U$ of the collapsing fluid can be obtained by taking variations of sectorial radius with respect to proper time.
\begin{equation}\label{24}
U=D_T C=\frac{\dot{C}}{A}.
\end{equation}
Using Eq.(19)-(22), with (24) we get from (23),

\begin{equation}\label{25}
\begin{split}
D_Tm &= -\frac{1}{2}\Bigg[U\Bigg\{(1+f_T)(\bar{P_r}-\frac{4}{3}\eta \sigma)+\mu f_T + \frac{\mathcal{G}}{2}\Bigg(\frac{f}{\mathcal{G}}-f_\mathcal{G}\Bigg)-\frac{\psi_{11}}{H^2}\Bigg\} \\
& +E\Bigg\{(1+f_T)\bar{q}+\frac{\psi_{01}}{AH}\Bigg\}\Bigg]C^2,
\end{split}
\end{equation}

\begin{equation}\label{26}
D_Cm= \frac{C^2}{2}\Bigg[\bar{\mu}+\varepsilon f_T-\frac{\mathcal{G}}{2}\Bigg(\frac{f}{\mathcal{G}}-f_\mathcal{G}\Bigg)-\frac{\psi_{00}}{A^2} +\frac{U}{E} \Bigg\{(1+f_T)\bar{q}+\frac{\psi_{01}}{AH}\Bigg\} \Bigg],
\end{equation}
where $\bar{\mu}=\mu+\varepsilon$, $\bar{P}_r=P_r+\varepsilon$, $\bar{q}=q+\varepsilon$ and $D_C=\frac{1}{C'}\frac{\partial}{\partial r}$. Now taking integral on both sides of Eq.(26) we obtain

\begin{equation}\label{27}
m= \int_{0}^{C}\frac{C^2}{2}\Bigg[\bar{\mu}+\varepsilon f_T -\frac{\mathcal{G}}{2}\Bigg(\frac{f}{\mathcal{G}}-f_\mathcal{G}\Bigg)- \frac{\psi_{00}}{A^2} +\frac{U}{E}\Bigg\{(1+f_T)\bar{q}+\frac{\psi_{01}}{AH}\Bigg\}\Bigg]dC,
\end{equation}
where $E\equiv\frac{C'}{H}$. Then Eq.(23) can be revised as

\begin{equation}\label{28}
E\equiv\frac{C'}{H}=\Bigg[U^2-\frac{2m}{C}+1 \Bigg].
\end{equation}
The particular combinations of $f(\mathcal{G}, T)$ corrections, structural variables and energy density via mass function can
be completed and is given as

\begin{equation}\label{29}
\frac{3m}{C^3}= \frac{3}{2C^3}  \int_{0}^{r}\Bigg[\bar{\mu}+\varepsilon f_T -\frac{\mathcal{G}}{2}\Bigg(\frac{f}{\mathcal{G}}-f_\mathcal{G}\Bigg)- \frac{\psi_{00}}{A^2} +\frac{U}{E}\Bigg\{(1+f_T)\bar{q}+\frac{\psi_{01}}{AH}\Bigg\}\Bigg]C^2C'dr,
\end{equation}

The usual components of Weyl tensor can be split in electric and magnetic parts but the magnetic
part of Weyl Tensor becomes zero due to spherical symmetry. So the Weyl tensor can be represented in terms of its electric part.
The electric part of Weyl tensor is defined as

\begin{equation}\nonumber
E_{\alpha \beta}=C_{\alpha\phi\beta\phi}V^\phi V^\varphi.
\end{equation}
The electric part of Weyl tensor can be rewritten as

\begin{equation}\nonumber
E_{\alpha\beta}=\varepsilon\Bigg[\chi_\alpha \chi_\beta - \frac{1}{3}(g_{\alpha\beta}+V_\alpha V_\beta)\Bigg],
\end{equation}
where $\varepsilon$ is the Weyl scalar and is given as

\begin{equation}\label{30}
\begin{split}
\varepsilon &= \frac{1}{2A^2}\Bigg[\frac{\ddot{C}}{C}-\frac{\ddot{H}}{H} - \Bigg(\frac{\dot{C}}{C} -\frac{\dot{H}}{H} \Bigg)\Bigg(\frac{\dot{A}}{A} +\frac{\dot{C}}{C} \Bigg)  \Bigg] -\frac{1}{2C^2} \\
& +\frac{1}{2H^2}\Bigg[\frac{A''}{A}-\frac{C''}{C} + \Bigg(\frac{H'}{H} +\frac{C'}{C} \Bigg)\Bigg(\frac{C'}{C} -\frac{A'}{A} \Bigg)   \Bigg].
\end{split}
\end{equation}
Using Eq.(19)-(22), (23) and (29) we can rewrite (30) as

\begin{equation}\label{31}
\begin{split}
\varepsilon &= \frac{1}{2}\Bigg[\bar{\mu} + \varepsilon f_T- (1+f_T)(\bar{\Pi}-2\eta\sigma)-\frac{\mathcal{G}}{2}\Bigg(\frac{f}{\mathcal{G}}-f_\mathcal{G}\Bigg)-\frac{\psi_{00}}{A^2} +\frac{\psi_{11}}{H^2} -\frac{\psi_{22}}{C^2}\Bigg] \\
& -\frac{3}{2C^3}  \int_{0}^{r}\Bigg[\bar{\mu}+\varepsilon f_T -\frac{\mathcal{G}}{2}\Bigg(\frac{f}{\mathcal{G}}-f_\mathcal{G}\Bigg)- \frac{\psi_{00}}{A^2} +\frac{U}{E}\Bigg\{(1+f_T)\bar{q}+\frac{\psi_{01}}{AH}\Bigg\}\Bigg]C^2C'dr,
\end{split}
\end{equation}
where $\bar{\Pi}=\bar{P}_r-P_{\perp}$.
Here the above expression connects density inhomogeneity and local anisotropy of pressure with Weyl tensor.

\section{Modified Scalar Variables and $f(\mathcal{G}, T)$ Gravity}
The $f(\mathcal{G}, T)$ gravity simply depends on the selection of $f(\mathcal{G}, T)$ model. For $f(\mathcal{G}, T)$ gravity one can have different models. We pick the following kind of model for $f(\mathcal{G}, T)$ gravity
\begin{equation}\label{32}
f(\mathcal{G},T)=f_1(\mathcal{G})+f_2(T),
\end{equation}
where $f_1(\mathcal{G})=\alpha \mathcal{G}^n$ given by Cognola et. al \cite{cogn1} and $f_2(T)=\lambda T$ with $\alpha$ and $\lambda$ as real numbers.

We present a couple of tensors defined by \cite{bel1}, \cite{herr2} and \cite{shar2}
\begin{equation}\nonumber
X_{\alpha \beta} = ^* R_{\alpha \gamma \beta \delta} ^{*} V^\gamma V^\delta = \frac{1}{2} \eta^{\varepsilon\rho}_{\alpha\gamma} R^{*}_{\varepsilon\rho\beta\delta} V^{\gamma} V^{\delta},
\end{equation}
where $R^*_{\alpha \beta \gamma \delta}=\frac{1}{2}\eta_{\varepsilon\rho\gamma\delta}R^{\varepsilon\rho}_{\alpha\beta}$
and
\begin{equation}\nonumber
Y_{\alpha \beta} =  R_{\alpha \gamma \beta \delta}  V^\gamma V^\delta.
\end{equation}
To develop the formalism for structure scalars in $f(\mathcal{G},T)$ gravity, we orthogonally split the Riemann curvature tensor. We develop

\begin{equation}\label{33}
\begin{split}
X_{\alpha \beta}&=X_{\alpha \beta}^{(m)}+X_{\alpha \beta}^{(D)} = \frac{1}{3}\Bigg[\bar{\mu}+\varepsilon f_T - \frac{\mathcal{G}}{2}\Bigg(\frac{f}{\mathcal{G}}-f_\mathcal{G}\Bigg) - \frac{\psi_{00}}{A^2} \Bigg]h_{\alpha \beta} \\
&- \frac{1}{2}\Bigg[(1+f_T)(\bar{\Pi}-2\eta\sigma)-\frac{\psi_{11}}{H^2}+\frac{\psi_{22}}{C^2}\Bigg]\Bigg(\chi_\alpha \chi_\beta - \frac{1}{3}h_{\alpha \beta} \Bigg)-E_{\alpha \beta},
\end{split}
\end{equation}

\begin{equation}\label{34}
\begin{split}
Y_{\alpha \beta}&=Y_{\alpha \beta}^{(m)}+Y_{\alpha \beta}^{(D)} = \frac{1}{6}\Bigg[\bar{\mu}+\varepsilon f_T +3\mu \lambda +(1+f_T)(3P_r - 2\bar{\Pi})- \frac{\psi_{00}}{A^2} - \frac{\psi_{11}}{H^2} \\ &- \frac{2\psi_{22}}{C^2}+
 \frac{\mathcal{G}}{2}\Bigg(\frac{f}{\mathcal{G}}-f_\mathcal{G}\Bigg)\Bigg] h_{\alpha \beta}
-\frac{1}{2}\Bigg[(1+f_T)(\bar{\Pi}-2\eta\sigma) - \frac{\psi_{11}}{H^2} + \frac{\psi_{22}}{C^2}\Bigg]\\ &\times \Bigg(\chi_\alpha \chi_\beta - \frac{1}{3}h_{\alpha \beta} \Bigg)+E_{\alpha \beta}.
\end{split}
\end{equation}
These tensors are composed of their trace and trace-less parts given as

\begin{eqnarray}
  X_{\alpha \beta} &=& \frac{1}{3}TrXh_{\alpha \beta}+X_{<\alpha \beta>}, \\
  Y_{\alpha \beta} &=& \frac{1}{3}TrYh_{\alpha \beta}+Y_{<\alpha \beta>},
\end{eqnarray}
where

\begin{eqnarray}
  X_{<\alpha \beta>} &=& h_\alpha^\rho h_\beta^\gamma \Bigg(X_{\rho \gamma} -\frac{1}{3} TrXh_{\rho \gamma}   \Bigg), \\
  Y_{<\alpha \beta>} &=& h_\alpha^\rho h_\beta^\gamma \Bigg(Y_{\rho \gamma} -\frac{1}{3} TrYh_{\rho \gamma}   \Bigg).
\end{eqnarray}
We can rewrite $ X_{<\alpha \beta>}$ and $ Y_{<\alpha \beta>}$ in another way
\begin{eqnarray}
  X_{<\alpha \beta>} &=& X_{TF} \Bigg(\chi_\alpha \chi_\beta - \frac{1}{3}h_{\alpha \beta} \Bigg), \\
  Y_{<\alpha \beta>} &=& Y_{TF} \Bigg(\chi_\alpha \chi_\beta - \frac{1}{3}h_{\alpha \beta} \Bigg).
\end{eqnarray}

Eq.(33) and (34) can be written separately as trace and trace-free components

\begin{equation}
TrX \equiv X_T = \bar{\mu}+\varepsilon \lambda- \frac{\alpha (1-n)}{2} \mathcal{G} ^n-\frac{\lambda}{2}T - \frac{\hat{\psi}_{00}}{A^2},
\end{equation}

\begin{equation}
\begin{split}
TrY &\equiv Y_T = \frac{1}{2}\Bigg[\bar{\mu}+\varepsilon\lambda + 3\mu \lambda + 3(1+\lambda)\bar{P_r}
-2(1+\lambda)\bar{\Pi} -\frac{\hat{\psi}_{00}}{A^2}-\frac{\hat{\psi}_{11}}{H^2} \\ & -\frac{2 \hat{\psi}_{22}}{C^2} +\alpha (1-n) \mathcal{G}+\lambda T \Bigg],
\end{split}
\end{equation}

\begin{eqnarray}
  X_{TF} &=& -\varepsilon -\frac{1}{2} \Bigg[(1+\lambda)(\bar{\Pi}-2\eta\sigma)-\frac{\hat{\psi}_{11}}{H^2}+\frac{ \hat{\psi}_{22}}{C^2}\Bigg], \\
  Y_{TF} &=& \varepsilon -\frac{1}{2} \Bigg[(1+\lambda)(\bar{\Pi}-2\eta\sigma)-\frac{\hat{\psi}_{11}}{H^2}+\frac{ \hat{\psi}_{22}}{C^2}\Bigg],
\end{eqnarray}
where the hat represents the dark source variables are computed after utilizing $f(\mathcal{G}, T)$ model. Using Eq.(29), (31) and (32) $X_{TF}$ and $Y_{TF}$ become

\begin{equation}
\begin{split}
X_{TF}&=-\frac{1}{2} \Bigg[\bar{\mu}+\varepsilon\lambda-\frac{\alpha (1-n)}{2}\mathcal{G}^n-\frac{\lambda}{2}T-\frac{\hat{\psi}_{00}}{A^2}\Bigg]+\frac{3}{2C^3}\int_{0}^{r}\Bigg[\bar{\mu}+\varepsilon\lambda \\ &-\frac{\alpha (1-n)}{2}\mathcal{G}^n-\frac{\lambda}{2}T-\frac{\hat{\psi}_{00}}{A^2}+\frac{U}{E}\Bigg\{(1+\lambda)\bar{q}+\frac{\hat{\psi}_{01}}{AH}\Bigg\}\Bigg]C^2C'dr,
\end{split}
\end{equation}

\begin{equation}
\begin{split}
Y_{TF}&=\frac{1}{2} \Bigg[\bar{\mu}+\varepsilon\lambda-2(1+\lambda)(\bar{\Pi}-2\eta\sigma)-\frac{\alpha (1-n)}{2}\mathcal{G}^n-\frac{\lambda}{2}T-\frac{\hat{\psi}_{00}}{A^2} \\ &+\frac{2\hat{\psi}_{11}}{H^2}-\frac{2\hat{\psi}_{22}}{C^2}\Bigg] -\frac{3}{2C^3}\int_{0}^{r}\Bigg[\bar{\mu}+\varepsilon\lambda-\frac{\alpha (1-n)}{2}\mathcal{G}^n-\frac{\lambda}{2}T-\frac{\hat{\psi}_{00}}{A^2} \\ &+\frac{U}{E}\Bigg\{(1+\lambda)\bar{q}+\frac{\hat{\psi}_{01}}{AH}\Bigg\}\Bigg]C^2C'dr.
\end{split}
\end{equation}

After using some effective fluid variables Eq.(41)-(44) can be written as
\begin{equation}
 \begin{split}
 X_{TF}&=-\frac{1}{2} \Bigg[\mu_{eff}+\varepsilon\lambda-\frac{\alpha (1-n)}{2}\mathcal{G}^n-\frac{\lambda}{2}T \Bigg] +\frac{3}{2C^3}\int_{0}^{r}\Bigg[\mu_{eff}+\varepsilon\lambda \\ &-\frac{\alpha (1-n)}{2}\mathcal{G}^n-\frac{\lambda}{2}T +\frac{U}{E}\Bigg\{(1+\lambda)\bar{q}+\frac{\hat{\psi}_{01}}{AH}\Bigg\}\Bigg]C^2C'dr,
 \end{split}
 \end{equation}

\begin{equation}
\begin{split}
Y_{TF}&=\frac{1}{2} \Bigg[\mu_{eff}+\varepsilon\lambda-\frac{\alpha (1-n)}{2}\mathcal{G}^n-\frac{\lambda}{2}T-2(1+\lambda)\Pi^{eff}+2\lambda\Bigg(\frac{\hat{\psi}_{22}}{C^2} \\ &-\frac{\hat{\psi}_{11}}{H^2} \Bigg) \Bigg] -\frac{3}{2C^3}\int_{0}^{r}\Bigg[\mu_{eff}+\varepsilon\lambda-\frac{\alpha (1-n)}{2}\mathcal{G}^n-\frac{\lambda}{2}T \\ &+\frac{U}{E}\Bigg\{(1+\lambda)\bar{q}+\frac{\hat{\psi}_{01}}{AH}\Bigg\}\Bigg]C^2C'dr,
\end{split}
\end{equation}

\begin{equation}
X_T = \mu_{eff}+\varepsilon \lambda- \frac{\alpha (1-n)}{2} \mathcal{G} ^n-\frac{\lambda}{2}T,
\end{equation}

\begin{equation}
\begin{split}
Y_T &= \frac{1}{2}\Bigg[\mu_{eff}(1+3\lambda)-2\varepsilon\lambda + 3(1+\lambda)P_r^{eff} - 2(1+\lambda)\Pi^{eff}
+\lambda \\ & \times\Bigg(3\frac{\hat{\psi}_{00}}{A^2}+\frac{\hat{\psi}_{11}}{H^2}+\frac{2 \hat{\psi}_{22}}{C^2}\Bigg) +\alpha (1-n) \mathcal{G}+\lambda T \Bigg],
\end{split}
\end{equation}
where $\mu_{eff}\equiv \bar{\mu}-\frac{\hat{\psi}_{00}}{A^2}$ ,
 $P_r^{eff}\equiv\bar{P_r}-\frac{\hat{\psi}_{11}}{H^2}-\frac{4}{3}\eta\sigma$,
 $P_\perp^{eff}\equiv P_\perp-\frac{\hat{\psi}_{22}}{C^2}+\frac{2}{3}\eta\sigma$ ,
 $\Pi^{eff}\equiv P_r^{eff}-P_\perp^{eff}= \Pi - 2\eta\sigma + \frac{\hat{\psi}_{22}}{C^2}-\frac{\hat{\psi}_{11}}{H^2}$ are effective fluid variables.

The dynamical development of relativistic compact frameworks have a resemblance with these structure scalars. $X_T$ has important significance in the definition of stellar power density coupled with terms of dark energy source. The evolution equation \cite{herr2} involving tidal forces and fluid parameter variables is
\begin{equation}
\begin{split}
& \Bigg[X_{TF}+\frac{1}{2}\Bigg\{\mu_{eff}+\varepsilon\lambda-\frac{\alpha (1-n)}{2} \mathcal{G} ^n-\frac{\lambda}{2}T\Bigg\}\Bigg]'=-X_{TF}\frac{3C'}{C} \\
& +\frac{\Theta-\sigma}{2}\Bigg[\frac{\hat{\psi}_{01}}{AH}+\bar{q}H(1+\lambda)\Bigg].
\end{split}
\end{equation}
The above expression shows that without dark source terms and radiating variables, we can have the following result
\begin{equation}\nonumber
\bar{\mu}'_{eff}=0\Leftrightarrow X_{TF}=0,
\end{equation}
which shows that $X_{TF}$ is an inhomogeneity factor. Other structure scalars describe expansion rate (Raychaudhuri) and shear evolution. The Raychaudhuri equation in our case is
\begin{equation}
-Y_T = V^\alpha \Theta_{;\alpha}+\frac{2}{3}\sigma^2+\frac{\Theta^2}{3}-a^{\alpha}_{;\alpha},
\end{equation}
and
\begin{equation}
Y_{TF}=a^2 + \chi^\alpha a_{;\alpha} -
\frac{aC'}{HC}-\frac{2}{3}\Theta\sigma-V^\alpha
\sigma_{;\alpha}-\frac{1}{3}\sigma^2.
\end{equation}

\section{Evolution Equations with Constant $\mathcal{G}$ and $T$}
In this section, we will discuss the modified structure scalars for the dust cloud with Gauss-Bonnet invariant and $T\equiv T^\alpha_\alpha$. The quantity of mass in this case is
\begin{equation}
m=\frac{1}{2}\int_{0}^{r}(\mu)C^2C'dr-\frac{\alpha(1-n)\tilde{\mathcal{G}}+\lambda \tilde{T}}{4}\int_{0}^{r}C^2C'dr,
\end{equation}
where tilde indicates that these terms are considered with reference to the constant backgrounds. After doing some calculations, the mass function and $\varepsilon$ for dust cloud become
\begin{equation}
\frac{3m}{C^3}=\frac{1}{2}\Bigg[\mu-\frac{1}{C^3}\int_{0}^{r}\mu'C^3dr\Bigg]-\frac{\alpha(1-n)\tilde{\mathcal{G}}+\lambda \tilde{T}}{2},
\end{equation}

\begin{equation}
\varepsilon = \frac{1}{2C^3}\int_{0}^{r}\mu'C^3dr-\frac{\alpha(1-n)\tilde{\mathcal{G}}+\lambda \tilde{T}}{4}.
\end{equation}

Eq. (55) and (56)are equivalent to Eq.(29) and (31). The structure scalars for dust cloud are
\begin{eqnarray}
  \tilde{X}_T &=& \mu - \frac{\alpha(1-n)\tilde{\mathcal{G}}}{2}-\frac{\lambda\tilde{T}}{2}, \\
  \tilde{Y}_T &=& \frac{1}{2}\Bigg[\mu+\alpha(1-n)\tilde{\mathcal{G}}+\lambda\tilde{T}\Bigg],
\end{eqnarray}

\begin{equation}
-\tilde{X}_{TF}=\tilde{Y}_{TF}=\varepsilon.
\end{equation}

The equations describing evolution and shear expansion become
\begin{eqnarray}
  -\tilde{Y}_T &=& V^\alpha \Theta_{;\alpha}+\frac{2}{3}\sigma^2+\frac{\Theta^2}{3}-a^{\alpha}_{;\alpha}, \\
  \tilde{Y}_{TF} &=& \varepsilon = -\frac{2}{3}\Theta\sigma-V^\alpha
\sigma_{;\alpha}-\frac{1}{3}\sigma^2.
\end{eqnarray}

The differential equation showing the inhomogeneity factor is
\begin{equation}
\Bigg[\tilde{X}_{TF}+\frac{1}{2}\mu-\frac{\alpha(1-n)\tilde{\mathcal{G}}+\lambda \tilde{T}}{4}\Bigg]'=-\tilde{X}_{TF}\frac{3C'}{C}
\end{equation}
from which it follows $\bar{\mu}'_{eff}=0\Leftrightarrow X_{TF}=0$ describing $X_{TF}$ as the inhomogeneity factor.

\section{Discussion}
In this paper, the dynamical system of compact objects has been discussed by taking $f(\mathcal{G}, T)$ gravity into consideration. We considered non-static spherically symmetric system coupled with anisotropic stresses radiating through heat flux and streaming approximations. After doing some fundamental calculations we have affiliated Weyl scalar with structural variables. Next we investigated the factors affecting the tidal forces in the development of collapsing spherical matter distribution in $f(\mathcal{G}, T)$ gravity. For this, we took a particular model of $f(\mathcal{G}, T)$ gravity which is given as $f(\mathcal{G},T)=f_1(\mathcal{G})+f_2(T)$. We have searched the role of $f(\mathcal{G}, T)$ terms given by dark energy source in the explanation of scalar functions. These scalars were calculated by orthogonally decomposing the Riemann curvature tensor. We have discovered that these scalar functions control the evolutionary systems in our celestial body.

Eq.(41) which is itself a structure scalar expresses the energy density together with the dark source variables in spherically symmetric dissipative anisotropic fluid distributions.

$Y_T$ comes out to be the mass density for dynamical system and this quantity is being controlled by means of anisotropic pressure together with dark source terms. The mass density is clearly linked with pressure anisotropy, radiating and non-radiating distributions along side $f(\mathcal{G}, T)$ gravity corrections. In Eq.(42) $Y_T$ has a connection with non-dissipative energy density.

For pressure anisotropy expansion free constraint is required. Eq.(52) and (60) shows that $Y_T$ controls the evolution of expansion scalar. As a result, $Y_T$ is significant to recognize the exposure of vacuum cavity inside the stellar object. It can be seen from Eq.(50) that $Y_T$ has a direct link with pressure anisotropy as well as $f(\mathcal{G}, T)$ corrections.

The impact of shear, local pressure anisotropy and tidal forces together with dark energy can be seen in Eq.(44). The shear evolution has been fully controlled by this scalar function $Y_{TF}$ also mentioned in Eq.(53). One needs to analyse the action of $Y_{TF}$ in order to interpret the role of shear on dynamical stages of radiating compact object.

$X_{TF}$ controls the energy density inhomogeneity for the anisotropic fluid as well as for dust perfect fluid \cite{herr3} but in Eq.(51) dissipative fluid parameters along with $f(\mathcal{G}, T)$ corrections cause an interruption in the contribution of $X_{TF}$. Although, if expansion becomes equivalent to shear, then due to presence of $f(\mathcal{G}, T)$ corrections some interruptions will appear in the development of inhomogeneity in relativistic systems. Hence, $X_{TF}$ controls the energy density irregularities with dark source variables.
In dust cloud with constant $\mathcal{G}$ and $T$, it is found that in $X_{TF}$ there are tidal forces that produced irregularities in the homogeneous stellar structures.

The above discussion proves that in modified gravity these structure scalars play an important role in the dynamics of self gravitating systems.

\end{document}